\documentclass{iopart}
\usepackage{setstack}
\usepackage{fancyhdr}
\usepackage{iopams}
\usepackage{amssymb}
\usepackage[dvips]{graphicx}
\usepackage{makeidx}
\newcommand{\one}{1}
\newcommand{\two}{2}
\newcommand{\three}{3}

\newcommand{\five}{4}
\newcommand{\six}{5}
\newcommand{\seven}{6}
\begin{document}
\title{Physical Observables for Noncommutative Landau Levels}
\author{Mauro Riccardi}
\address{Dipartimento di Fisica, Universit\`a di Firenze, \\ via G.Sansone 1, I-50019 Sesto Fiorentino, Italy}
\ead{riccardi@fi.infn.it}
\newcommand{\bra}[1]{ \langle #1 |}
\newcommand{\ket}[1]{ | #1 \rangle}
\newcommand{\braket}[2]{\langle #1 | #1 \rangle}
\newcommand{\binomial}[2]{\left(\begin{array}{c} #1 \\[1pt] #2 \end{array}\right)}
\newcommand{\twoo}{\!\!\begin{array}{c} \scriptscriptstyle{o} \\[-8pt] \scriptscriptstyle{o} \end{array}\!\!}
\newcommand{\weyl}[1]{\twoo #1 \twoo}
\newcommand{\weq}{\approx}		 
\newcommand{\throw}{\longrightarrow}
\newcommand{\mapto}{\mapsto}
\newcommand{\implies}{\Longrightarrow}
\newcommand{\implied}{\Longleftarrow}
\renewcommand{\mapsto}{\longmapsto}
\newcommand{\sfrac}[2]{#1\diagup #2}	   
\newcommand{\scalprod}[2]{\bigl( #1 \, \vert \, #2\bigr)} 
\newcommand{\pref}[1]{(\ref{#1})}        
\newcommand{\C}{\mathbb C}
\newcommand{\R}{\mathbb R}
\newcommand{\foreach}{\forall}		 
\newcommand{\hbz}{\hat{\bar z}}

\begin{abstract}
The Quantum Mechanics of a point particle on a Noncommutative Plane in a magnetic field is implemented in the present work as a deformation of the algebra which defines the Landau levels.
I show how to define, in this deformed Quantum Mechanics, the physical observables, like the density correlation functions and Green function, on the completely filled ground level. Also it will be shown that the deformation changes the effective magnetic field which acts on the particles at long range, leading to an incompressible fluid with fractional filling of Laughlin type.
\end{abstract}
\pacs{02.40.Gh,03.65.-w}
\section{Introduction}
The aim of the present paper is to analyse, in physical terms, what the introduction of the noncommutativity in a geometry can imply on a physical system, and to study the resulting theory in order to see whether it can describe physical objects as well. Following Nair and Polychronakos (see \cite{qmonncplandsph}), we consider the Noncommutative Plane in presence of a constant magnetic field, leading to Noncommutative Landau levels. The techniques employed are as close as it is possible to those of standard Quantum Mechanics computations. \\
In section 2 we will briefly review some standard facts regarding the quantum mechanics of charged particles on a plane in the presence of a constant magnetic field, in particular some properties of the Landau levels, second quantization, and the role of $W_\infty$ algebra in the study of incompressibility of the fluid of electrons in the lowest Landau level.\\
In section 3 we will generalise the geometry of the plane by rescaling the flux of the magnetic field $\mathbf B$ through a unit area element. We will show in detail that this deformation, modifying the magnetic translations, affects the geometry as well, rendering it noncommutative in a very natural way. Then we will show the procedure to define and compute the correlation functions of the density on the incompressible ground state. The one and two points correlation functions will be computed, and the first physical consequences of noncommutativity will be shown. The short distance and long distance behaviours of the fluid will be taken into consideration, the former being related with the delocalization induced by noncommutativity, and the latter giving information about the filling fraction of the fluid itself.\\
In section 4 we will propose a definition for the Green function in this framework, and compute it in the simplest case. \\
We observe that the results in the sections 3 and 4 are new, as far as we know. The present paper originated within the bounds of the study of Quantum Hall effect, keeping in mind Susskind's proposal \cite{QH-NCCS}: yet it belongs to the more general context of Quantum Mechanics (e.g. see \cite{qmonncplandsph}). Hence the aim of the paper is two-fold: by one hand to discuss in some detail the physics of noncommutativity, and by the other hand to find and test a useful set of tools to describe the degrees of freedom of the fluid of two-dimensional electrons.

\section{Landau levels}
I will start recalling the Quantum Mechanics of planar particles in a magnetic field, including a discussion about the so called $W_\infty$-algebra which is important to describe incompressibility. 
\subsection{The one body problem}
The hamiltonian for an electron in a uniform constant
magnetic field may be written as
\begin{equation}
\mathfrak H = \frac{1}{2m} \left(\mathbf p-\frac{e}{c}\mathbf A\right)^2 \ ,
\end{equation}
whereas the potential in the symmetric gauge is
\[
\mathbf A = \frac{B}{2} ( -x_2 , x_1 ) \ .
\]
It is also customary to use {\em magnetic units}, defined by:
\[
\hslash = 1 , \quad c = 1 , \quad \ell = \sqrt{\frac{2\hslash c}{e B}} = 1 \ .
\]
The quantity {\em magnetic length} $\ell$ is the scale of the problem introduced by the
presence of the magnetic field. \\ 
The hamiltonian may be written in a harmonic oscillator form, introducing the ladder operators\footnote{The $\hat a$ and $\hat a^\dagger$ operators are manifestly the covariant derivatives.}
\begin{equation}
\hat a \doteq \frac{z}{2} + \bar\partial \qquad \hat a^\dagger \doteq \frac{\bar z}{2} - \partial \ .
\end{equation}
They satisfy the usual commutation relation
\begin{equation}
\left[ \hat a \, , \, \hat a^\dagger \right] = 1 \ .
\end{equation}
So the hamiltonian takes the form
\begin{equation}
\mathfrak H = 2 \ \hat a^\dagger \, \hat a \, + \, 1 \ .
\label{eqn:oscillatorhamiltonian}
\end{equation}
There is another conserved quantity, the angular momentum, which is conserved due to the rotational invariance. To write it, we introduce two more ladder operators, commuting with the $\hat a$'s
\begin{equation}
\hat b \doteq \frac{\bar z}{2} + \partial \qquad \hat b^\dagger \doteq \frac{z}{2} -\bar\partial \ .
\end{equation}
They satisfy the equation
\begin{equation}
\bigl[ \hat b \, , \, \hat b^\dagger \bigr] = 1 \ .
\label{eqn:bopscom}
\end{equation}
These operators are the generators of the group of magnetic translations, whose elements are translations {\em plus} gauge transformations compensating for the variation of $\mathbf A$ under the translation itself\cite{fubini,fradkinbook}. 
In ordinary units the commutator of $b$ operators \pref{eqn:bopscom} would be
\[
\bigl[ \hat b,\hat b^\dagger \bigr]=\frac{B}{2} \ .
\]
The angular momentum can be written as
\begin{equation}
\mathfrak J = \ \hat b^\dagger \, \hat b \, - \, a^\dagger \, a \ .
\label{eqn:angularmom}
\end{equation}
We see that $[\mathfrak H,\mathfrak J]=0$ so that a base for the Hilbert space is given in term of simultaneous eigenstates of both the operators, in this form \index{Landau levels!base of states}
\begin{equation}
\left\{
\begin{array}{ccl}
\mathfrak H \psi_{mn} & = & (2n+1) \ \psi_{mn} \\
\mathfrak J \psi_{mn} & = & (m-n) \ \psi_{mn}
\end{array}\right. \quad 
\implies \quad
\psi_{mn} \doteq \frac{\hat b^{\dagger m}}{\sqrt{m!}}\frac{\hat a^{\dagger n}}{\sqrt{n!}} \ \psi_0
\label{eqn:landaulevels}
\end{equation}
The states $\psi_{mn}$ are normalised by
\[
\langle \psi_{mn} | \psi_{kl}\rangle = \int \rmd^2z \, \psi_{mn}^\ast(z,\bar z)\psi_{kl}(z,\bar z) \rme^{-|z|^2} \ .
\]
The basic wave-function $\psi_0(z,\bar z)=\langle z,\bar z \, | \, \psi_0\rangle$ is the gaussian
\begin{equation}
\langle z,\bar z \, | \, \psi_0\rangle = \psi_0(z,\bar z) = \frac{1}{\sqrt\pi} \ \rme^{-\frac{|z|}{2}^2} \ , \quad 
\left\{\begin{array}{ccl}
\hat a \ket{\psi_0} & = & 0 \\
\hat b \ket{\psi_0} & = & 0 
\end{array}\right.
\label{eqn:gausswf}
\end{equation}
Each energy level (Landau levels) is infinitely degenerate. The wavefunctions of the states in the level $n=0$ (the {\em lowest Landau level}), are
\begin{equation}
\psi_{m0}(z,\bar z) = \frac{1}{\sqrt\pi}\frac{z^m}{\sqrt{m!}} \ \rme^{-\frac{|z|}{2}^2} \ .
\label{eqn:lowestwf}
\end{equation}

These are the wave functions of particles localised in a ``fuzzy''
annulus, because the probability distribution is
angle-independent and peaked at $|z|^2=m$. So the lowest level is made
up by concentric layers. In the higher Landau levels, the wave functions present,
besides the power factor, a generalised Laguerre polynomial factor.

We may count the states in each Landau level, in a disc
of radius $R$, their number being
$n_e=\frac{R^2}{\ell^2}=\frac{\Phi}{\Phi_0}$ being $\Phi = \pi R^2 B$
the magnetic flux through the disc and $\Phi_0=\pi\ell^2B$ the {\em
quantum of magnetic flux}\index{Magnetic flux!elementary fluxon}. So
we may say that in each Landau level there is one state for each {\em
flux quantum} through the disc\index{Magnetic flux!flux quanta}.
\subsection{$W_\infty$ algebra}\index{Incompressibility!$W_\infty$ algebra}\label{sec:winf}
By using the fact that the generators of magnetic translation $\hat b$, $\hat b^\dagger$ commute with the hamiltonian $\mathfrak H$, we can construct infinite conserved quantities \cite{infsym}
\begin{equation}
\mathcal L_{nm} \doteq (\hat b^\dagger)^n \hat b^m \quad , \ n,m\in \mathbb N \ .
\label{eqn:winfgen}
\end{equation}
We may ask now which N\"other symmetry they generate. Their algebra is (here and in the following $m\curlywedge k \doteq {\rm min}(m,k)$)
\begin{equation}
[\mathcal L_{nm},\mathcal L_{kl}] = \sum_{i=1}^{m\curlywedge k} \frac{m!k!}{(m-i)!(k-i)!i!}\,\mathcal L_{n+k-i,m+l-i} - \left(\begin{array}{c} m\leftrightarrow l \\ n\leftrightarrow k\end{array}\right)
\label{eqn:lmnqalg}
\end{equation}
which, up to higher quantum corrections (we restore for a moment $\hslash$), reads (see \cite{infsym})
\begin{equation}
[\mathcal L_{nm},\mathcal L_{kl}] = \hslash (mk-nl) \mathcal L_{n+k-1,m+l-1} + O(\hslash^2) \ .
\label{eqn:lmncalg}
\end{equation}
This is known to be the algebra of (classical) area preserving diffeomorphisms, or $w_\infty$. The algebra defined by \pref{eqn:lmnqalg}, like all the quantum generalisations of \pref{eqn:lmncalg}, is called $W_\infty$ algebra.
\subsection{Second quantization}
We can give now a physical interpretation of the generators of $W_\infty$ algebra (see \cite{infsym} and refs. therein) by using second quantization. Namely, given the wavefunctions \pref{eqn:landaulevels}, we define the field operators\footnote{Operators acting on Fock space will be marked by a check above them to distinguish them from first quantized operators, that are marked by a hat, following tradition.}
\[
\check \phi(z,\bar z) \doteq \sum_{ln} \check c_{ln} \psi_{ln}(z,\bar z) \ ,
\]
that include the fermionic Fock operators
\[
[\check c_{ln},\check c^\dagger_{km}]_+ = \delta_{lk}\delta_{nm} 
\]
acting on a Hilbert space defined starting from the Fock vacuum $\ket 0$.
The second quantized version of the $\mathcal L_{st}$ operators is
\begin{equation}
\mathcal L_{st} \doteq \int \rmd^2z \ \check \phi^\dagger(z,\bar z) \ (\hat b^\dagger)^s \, \hat b^t \ \check \phi(z,\bar z) = \sum_{n=0}^\infty \ \sum_{l=t}^\infty \check c^\dagger_{l,n}\check c_{s+l-t,n} \frac{\sqrt{l!(s+l-t)!}}{(l-t)!} \ .
\label{eqn:secondlst}
\end{equation}
Notice that the different Landau levels labelled by the number of $a^\dagger$.s in the state are not connected by the $\mathcal L_{st}$ operators. Each term of the sum \pref{eqn:secondlst}
simply shuffles the particles within the same ($n$-th) level, varying their angular momentum $l$ by $l\mapto l+s-t$. \\
Let us now look to the lowest Landau level ($n=0$), and to the action of $\mathcal L_{st}$ on the ground state. The latter is the state with the minimum angular moment, which is simply, for $N$ particles
\begin{equation}
\ket\Omega \doteq \check c^\dagger_{N,0}\cdots \check c^\dagger_{0,0}\ket 0 \ .
\label{eqn:gsdef}
\end{equation}
Applying a generator of $W_\infty$ to $\ket\Omega$, we notice immediately that it vanishes identically if $s<t$, while it reduces to a number in the case $s=t$
\begin{displaymath}
\left\{
\begin{array}{ll}
\mathcal L_{st}\ket\Omega = \ 0  & \Leftarrow \ t>s\\
\mathcal L_{ss}\ket\Omega = \frac{(N+1)!}{(s+1)(N-s)!} \ket\Omega &
\end{array}
\right. \ .
\end{displaymath}
So the only nontrivial case is when $s>t$, in which case its effect on the ground state is that of increasing the angular momentum of the ground state $\ket\Omega$ by shifting electrons from inside the {\em Fermi sphere} to more external orbitals. So the incompressibility of the ground state is simply due to the fact it is the state with minimum angular momentum, and can be written by the {\em highest weight-like} conditions (\cite{infsym})
\begin{equation}
\mathcal L_{st} \ket\Omega = 0 \ \ \Leftarrow \ s<t \ .
\label{eqn:highestw}
\end{equation}
We stress here that the commutation relations close within the set of $\mathcal L_{st}$ with $s<t$. So the whole Lie sub-algebra generated by $\{\mathcal L_{st}\}_{s<t}$ annihilates the ground state. In the next section it will be shown that incompressibility is not spoilt by the introduction of noncommutativity.

\section{Deformed Landau levels}
We start from the works \cite{qmonncplandsph} about quantum mechanics on noncommutative plane, and introduce effects of a noncommutative geometry. The procedure here will be slightly different by that of references \cite{qmonncplandsph}, in that we will generalise the algebra of the ladder operators $\hat a,\hat a^\dagger$ and $\hat b,\hat b^\dagger$, and impose some conditions on the quantum coordinate operators. We will find soon that in this process the plane becomes noncommutative. Also we will be able to compute second quantized physical quantities for the incompressible fluid, which was not in the scope of \cite{qmonncplandsph}.\\
The generalised algebra is
\begin{equation}
\label{eqn:deform}
\left\{
\begin{array}{ccl}
\left[ \hat a,\hat a^\dagger\right] & = & 1 \\[1pt]
\bigl[ \hat b,\hat b^\dagger \bigr] & = & \beta \ \in\mathbb R_0^+\\[1pt]
\bigl[\, \hat a,\hat b \,\bigr] & = & 0 \ = \ \bigl[ \hat a,\hat b^\dagger \bigr]\\
\end{array}\right. \ .
\end{equation}
We want to keep the interpretation of this algebra as that of the quantum mechanics on a plane thread by the magnetic field; therefore we take the $\hat a,\hat a^\dagger$ operators as the kinematic momenta with which the Hamilton operator is built up, and the $\hat b,\hat b^\dagger$ as the magnetic translations on the plane. The deformation of the commutator $[\hat b,\hat b^\dagger]$ implies that the flux of the magnetic field $\mathbf B$ through an unit area is rescaled by $\beta$. So we have
\[
\mathfrak H =2 \hat a^\dagger \hat a + 1 \qquad [\hat b,\mathfrak H]=0=[\hat b^\dagger,\mathfrak H] \ .
\]
We still have an Hilbert space built starting from a vacuum $\ket{\psi_0}$, by the application of both $\hat a$ and $\hat b$. We use the same notation we employed before in the ``ordinary'' case, see \pref{eqn:landaulevels}.
 
We can fix the form of the coordinate operators in terms of the $\hat a$'s and $\hat b$'s by considering what the commutation relations of the latter with $\hat z,\hbz$ must be. We have the requirements\footnote{These conditions respect the structure of Bargmann space of analytic functions for the lowest Landau level.}
\[
[\hat z,\hat a] = 0  \ , \ [\hat z,\hat a^\dagger] = 1
\]
just as in the ordinary case, and 
\[
[\hat b^\dagger,\hat z]=0 \ , \ [\hat b,\hat z] = 1 \ ,
\]
because of the transformation rules of the coordinates under magnetic translations. These relations fix the coordinates $\hat z,\hbz$ to be
\begin{equation}
\label{eqn:complcoords}
\left\{
\begin{array}{ccl}
\hat z&\doteq &\sfrac{\hat b^\dagger}{\beta} + \hat a\\
\hbz &\doteq &\sfrac{\hat b}{\beta} + \hat a^\dagger\\
\end{array}\right. \ .
\end{equation}
In order to keep the rotation invariance in our problem, we must fix the form of the angular momentum, $\mathfrak J$, such that it both commutes $\mathfrak H$, and transforms the coordinates in the natural (vector) fashion, i.e.
\[
\bigl[\mathfrak J,\mathfrak H\bigr] = 0 \qquad \bigl[\mathfrak J,\hat z\bigr] = \hat z \qquad \bigl[\mathfrak J,\hbz\bigr] = - \hbz \ .
\]
With this properties, $\mathfrak J$ is found to be
\[
\mathfrak J = \frac{\hat b^\dagger \hat b}{\beta}-\hat a^\dagger \hat a \ .
\]
Of course the normalised eigenvectors of $\mathfrak H$ and $\mathfrak J$ are modified in the following way
\begin{equation}
\psi_{mn} \doteq \frac{\hat b^{\dagger m}}{\sqrt{m!\beta^m}}\frac{\hat a^{\dagger n}}{\sqrt{n!}} \ \psi_0 \ . 
\label{eqn:betalandaulevels}
\end{equation}
As a consequence, the generators \pref{eqn:winfgen} of $W_\infty$ are generalized to be
\begin{equation}
\mathcal L_{nm} \doteq \left(\frac{\hat b^\dagger}{\sqrt{|\beta|}}\right)^n \left(\frac{\hat b}{\sqrt{|\beta|}}\right)^m \ .
\label{eqn:winfgengen}
\end{equation}
From the \pref{eqn:complcoords} the noncommutativity relation of the coordinates can be computed to be
\begin{equation}
[\hbz,\hat z] = \frac{1}{\beta} - 1 \ .
\label{eqn:mycommutator}
\end{equation}
Of course when $\beta=1$ the original commutative theory is
recovered. When $\beta\neq 1$, these coordinates do not have a straightforward meaning,
because they are not c-numbers: let us discuss this point in more detail. 
In the study the quantum mechanics
of a point charge in ordinary Landau levels, usually what one does is to pick up a pair of
functions from $\mathcal A=\mathbf C^{r\geq 0}(\mathbb R^2)$ (since we are on a plane), and identify
any value of the pair of coordinates, with a point on the
plane. In the quantum theory, there exists the position operator, and each
point of the plane corresponds to a vector in an orthonormal complete
set $\{\ket{z,\bar z}\}$ of eigenstates of position operator. 
In the more abstract algebraic framework (e.g. see \cite{landincbook}), a point on a space is basically an
equivalence class of irreducible representations of the algebra
$\mathcal A$ of ($\mathbf C^{r\ge 0}$) functions on that
space. From the same point of view of the above lines, each one of these equivalence classes is labelled by the eigenvalues of the coordinate operator, which are just c-numbers. The operators \pref{eqn:complcoords}, do not form a complete system of operators, because they cannot be simultaneously diagonalised, and do not lead to pairs of coordinates. Hence, from the coordinates, one obtains a less detailed information on the configuration of the system.

\subsection{The Weyl transform}
\label{sec:weyltransform}
A well-known method to deal with quantities depending on noncommutative coordinates is to consider Wigner functions. In this subsection we will see definitions which are customary in the study of noncommutative plane. Basically we want to study the matrix element\footnote{The fact that the we compute lowest Landau level matrix elements is due to the ground state we will study in the following. It does not imply any projection of the operators: the only source of noncommutativity is the algebraic deformation.}
\[
\scalprod{\psi_{l,0}}{\weyl{\delta(p-\hbz)\delta(q-\hat z)}\psi_{m,0}}
\]
between two one-particle states of the lowest Landau level; here $\weyl{\cdots}$ means we are taking the symmetric (Weyl) ordering, that avoids ambiguities in the definition of the above equation. Another reason is the following. Let us introduce now the Weyl transform which maps functions to operators. Take the algebra of functions on the plane, and take the algebra (noncommutative plane) $\mathcal A_{\theta}$ generated by the operators $\hat x^i$ satisfying
\[
\left[\hat x^i,\hat x^j\right] = \theta \epsilon^{ij},\quad i,j=1,2 \ .
\]
This relation is used as a starting point in many papers about Quantum Field Theory on noncommutative spaces. In the present paper we have instead the relation \pref{eqn:mycommutator}, which is a consequence of the deformation \pref{eqn:deform} and of the conditions we imposed on the coordinate operators in the previous subsection. 
We can associate to each function $f:\mathbb R^2\throw\C$ on the plane the operator of $\mathcal A_\theta$
\begin{equation}
\label{eqn:weyltransform}
U[f] \doteq \frac{1}{(2\pi)^2} \int \rmd^2 k \,\int d^2 \xi\, \rme^{i k\cdot (\hat x-\xi)} f(\xi) \ .
\end{equation}
This is a ``noncommutative generalisation'' of the Dirac $\delta$ relation
\[
f(x) = \int \rmd y\,\delta(x-y) f(y) \ ,
\]
so that we can write, using complex coordinates\footnote{We can observe that introducing this kind of Dirac $\delta$-functions in an operatorial expression is equivalent to use the collective coordinates as in \cite{spallucci}\label{collect}}:
\begin{equation}
U[f] = \int \, \rmd^2\xi \ f(\xi) \ \weyl{\delta(\xi_z - \hat z)\delta(\xi_{\bar z}- \hbz)} \ .
\label{eqn:weyltran}
\end{equation}
This formula gives a precise meaning to the idea of ``substituting'' an operator for a coordinate in an ordinary function; indeed it allows us to write each operator of $\mathcal A_\theta$ in an unambiguous form. Moreover the equation \pref{eqn:weyltransform} does automatically the job of ordering operator monomials in the most symmetric way. 
As is also well-known, the product of two operators is expressed by the Weyl transform in terms of the Moyal product.
By plugging in the definition of the Weyl operators, and using the Campbell-Baker-Hausdorff formula, we find
\begin{equation}
\label{eqn:moyalprod}
U[f]U[g] = \frac{1}{(2\pi)^2} \int \rmd^2 k \, \rme^{\rmi k\cdot \hat x} \int \rmd^2\xi\, \rme^{-\rmi k\cdot\xi} f\star g (\xi) = U[f\star g] \ ,
\end{equation}
where the Moyal product $\star$ is 
\[
f\star g (\xi) \doteq f(\xi) \ \rme^{\frac{\theta}{2} \overset{\leftarrow}{\partial_i} \; \epsilon^{ij} \; \overset{\rightarrow}{\partial_j}} \ g(\xi) \ .
\]
Every operatorial ordering of \pref{eqn:weyltran} defines a different quantization of the algebra of regular functions on the plane, but all of these quantizations are equivalent. Thus, we are free to chose the symmetric ordering, being the most natural one. 
The expression of the matrix element is:
\begin{eqnarray}
\label{eqn:matel1}
\fl\scalprod{\psi_{l,0}}{\weyl{\delta(q-\hat z)\delta(p-\hbz)}\psi_{m,0}} & \doteq \int \frac{\rmd x\rmd y}{(2\pi)^2}\, \scalprod{\psi_{l,0}}{\rme^{\rmi(qx+py)-\rmi(\hat z x+\hbz y)}\psi_{m,0}} = \nonumber \\
 & = \int \frac{\rmd x\rmd y}{(2\pi)^2}\, \rme^{\rmi(qx+py)} \scalprod{\rme^{\frac{\rmi}{\beta}\bar y\hat b^\dagger}\psi_{l,0}}{\rme^{-\frac{\rmi}{\beta}x\hat b^\dagger}\psi_{m,0}} \rme^{(\frac{xy}{2\beta}-\frac{xy}{2})} \, .
\end{eqnarray}
Here we used the fact that in the lowest Landau level the $\hat a$ operator vanishes, $\hat a\psi_{l,0}=0$. 
Now the computation leads for the matrix elements
\begin{eqnarray}
\scalprod{e^{\frac{i}{\beta}\bar y\hat b^\dagger}\psi_{l,0}}{e^{-\frac{i}{\beta}x\hat b^\dagger}\psi_{m,0}} = \sqrt{\frac{l!m!}{\beta^{l+m}}} \ \sum_{s=0}^{m\curlywedge l}\frac{\beta^s (-ix)^{l-s}(-iy)^{m-s}}{(l-s)!(m-s)!s!} \ e^{-\frac{xy}{\beta}} \ ,
\label{eqn:shortmatel}
\end{eqnarray}
where $m\curlywedge l \doteq\min{\{m,l\}}$.
Notice that this is just a polynomial in $x$ and $y$ times the overall exponential. Now we can put it back into \pref{eqn:matel1} and take the Fourier transform obtaining
\begin{eqnarray}
\fl\scalprod{\psi_{l,0}}{\weyl{\delta(q-z)\delta(p-\bar z)}\psi_{m,0}} = & \nonumber\\
= \frac{1}{\pi}\left|\frac{2\beta }{1+\beta}\right| \sqrt{\frac{l!m!}{\beta^{l+m}}} \ \sum_{s=0}^{m\curlywedge l} & \frac{\beta^s}{(l-s)!(m-s)!s!} \! \left(-\frac{\partial}{\partial q}\right)^{l-s}\!\left(-\frac{\partial}{\partial p}\right)^{m-s} e^{-\frac{2\beta}{1+\beta}pq} \, .
\label{eqn:deltas}
\end{eqnarray} 
The above formula allow us to write any expectation value of the form $\scalprod{\psi_{l,0}}{U[f] \ \psi_{m,0}}$ as an integral on a ``quasi-classical phase space'' $\{(q,p)\}$
\begin{eqnarray}
\fl\scalprod{\psi_{l,0}}{U[f] \ \psi_{m,0}} = \frac{\sqrt{l!m!}}{\pi}\frac{|2\beta |}{|1+\beta|} \, \sum_{s=0}^{m\curlywedge l}\sum_{t=0}^{(m\curlywedge l)-s} & \frac{(-1)^{t}\beta^{s-\frac{l+m}{2}}\left(\frac{2\beta}{1+\beta} \right)^{m+l-2s-t}}{(l-s-t)!\,(m-s-t)!\,s!\,t!} \ \times \nonumber \\
& \times \int \: \rmd q \, \rmd p \ f(q,p) \ \rme^{-\frac{2\beta}{1+\beta}pq} \  \, p^{l-s-t} \ q^{m-s-t} \, .
\label{eqn:weyltransfmatel}
\end{eqnarray}
Considering the expression \pref{eqn:lowestwf}, we can rescale the integral and recognise it as the matrix element between the wavefunctions of appropriate states in the lowest Landau level for $\beta=1$ so that we can write
\begin{equation}
\fl\scalprod{\psi_{l,0}}{U[f] \ \psi_{m,0}} = \sqrt{l!m!}\,\sum_{s=0}^{l\curlywedge m}\sum_{t=0}^{(l\curlywedge m)-s} \frac{(-1)^{t} \ \left(\frac{1+\beta}{2}\right)^{s-\frac{l+m}{2}}}{s!\, t!\sqrt{(m-s-t)!(l-s-t)!}} \ \scalprod{\psi^{\beta=1}_{l-s-t,0}}{\tilde f \, \psi^{\beta=1}_{m-s-t,0}} \, ,
\label{eqn:finalmatel}
\end{equation}
where
\[
\tilde f(\zeta,\bar\zeta) \doteq f\left(\sqrt{\frac{1+\beta}{2\beta}} \ \zeta,\sqrt{\frac{1+\beta}{2\beta}} \ \bar\zeta\right) \ .
\]
We see that \pref{eqn:weyltransfmatel} has been rewritten as a linear combination of the analogous matrix elements for $\beta=1$ involving all and only the states with lower angular momentum ($\psi_{l',0}$ with $l'\leq m\curlywedge l$). Notice that the algebraic deformation \pref{eqn:deform} does not reduce to a simple rescaling of the coordinates of $\beta=1$. The introduction of noncommutativity has instead deeper physical consequences, which require the study of some correlation functions. This will be done in the following.
Eqn. \pref{eqn:finalmatel} also implies that the deformation of the algebra considered here, does not violate the incompressibility defined in terms of $W_\infty$ algebra (see section \pref{sec:winf}): the matrix elements of any observables are indeed written in terms of $\beta=1$ matrix elements between states of equal or lower angular momentum. 
Since the deformed (rescaled) $W_\infty$ generators \pref{eqn:winfgengen} obey the same algebra \pref{eqn:lmnqalg} for each $\beta$, the ``noncommutative'' fluid described by \pref{eqn:gsdef} we are going to study is still incompressible in the ordinary sense (see \cite{infsym}).
\subsection{Density and its correlation functions}
Now we come back for a moment to the $\beta=1$ situation, i.e. to the commutative case, for the theory projected to the first Landau level. In this context, one has the wavefunctions \pref{eqn:lowestwf}, and the ground state of the incompressible fluid of $N+1$ electrons is still given by \pref{eqn:gsdef}. \\
Now we want to evaluate the expectation value of the density operator $\mathbf\rho$ of the field $\mathbf\phi$ on this fundamental state $\ket\Omega$. The density is
\[
\check \rho(z,\bar z) \doteq \check \phi^\dagger\check \phi(z,\bar z) = \sum_{kl} \, \check c^\dagger_l\check c_k\, \psi^\ast_{l,0}(z,\bar z)\psi_{k,0}(z,\bar z) \ .
\]
For its expectation value one finds
\begin{eqnarray}
\bra\Omega\check \rho(z,\bar z)\ket\Omega & \doteq \sum_{kl} \, \psi^\ast_{l,0}(z,\bar z)\psi_{k,0}(z,\bar z)\,\bra 0 \check c_0\cdots \check c_N \, \check c^\dagger_l\check c_k\,  \check c^\dagger_N\cdots \check c^\dagger_0\ket{0} & = \nonumber \\
& = \sum_{l=0}^N\, \psi^\ast_{l,0}(z,\bar z)\psi_{l,0}(z,\bar z) \ .
\end{eqnarray}
This can be written as
\[
\bra\Omega\check \rho(z,\bar z)\ket\Omega = \sum_{l=0}^N\, \int \rmd^2\zeta \ \psi_{l,0}(\zeta,\bar \zeta) \,\delta(\zeta-z)\delta(\bar\zeta-\bar z)\,\psi_{l,0}(\zeta,\bar\zeta) = \sum_{l=0}^N\, \scalprod{\psi_{l,0}}{\delta_z\delta_{\bar z}\psi_{l,0}} \ .
\]

For $\beta\neq 1$, we repeat the previous steps, obtaining the following relation
\[
{}_{\beta}{\bra\Omega U[\check \rho(\eta,\bar\eta)]\ket\Omega}_{\beta} = \sum_{k=0}^N\, \scalprod{\psi_{k,0}}{\weyl{\delta(\eta-\hat z)\delta(\bar\eta-\hbz)}\psi_{k,0}} \ ,
\]
where $\eta$ is a complex number which represents the point where we computed the density in the $\beta=1$ framework of above. Notice that in the above formula the Weyl transform of $\rho$ is consistent with its definition in \pref{eqn:weyltransform}, even if $\rho$ is an operator on Fock space. The definition \pref{eqn:weyltransform} indeed is trivially extended to act on functions of
\[
\mathcal F =\left\{\check f:\R^2\throw\C\otimes \mathcal O\right\} \ ,
\]
i.e. complex functions on the plane, tensored with the space of the operators on Fock space, since the coordinate operators act as a multiplication on states of the Fock space of the electrons. We can apply now our formula \pref{eqn:weyltransfmatel} to get the result after some manipulation
\begin{eqnarray}
\fl {}_{\beta}{\bra\Omega U[\check \rho(\eta,\bar\eta)]\ket\Omega}_{\beta} = \frac{1}{\pi}\left|\frac{2\beta}{1+\beta}\right|\sum_{k=0}^N\sum_{s=0}^k\binomial{k}{s}\left(\frac{2}{1+\beta}\right)^{k-s} \, \frac{\mathcal U\left(s-k,1,\left|\frac{2\beta}{1+\beta}\right|\eta\bar\eta\right)}{(k-s)!} \ \rme^{-\left|\frac{2\beta}{1+\beta}\right|\eta\bar\eta} \, ,
\end{eqnarray}
where $\mathcal U(a,c,z)$ is the Tricomi function (hypergeometric confluent of the second kind, see \cite{handbook}).

We can now put the complex coordinates of any point in the place of $\eta$ and $\bar\eta$, so that we can see that the expectation value of the density on the lowest Landau level is rotational invariant. We can plot it for various values of $\beta$ and at fixed $N$ (see figure \one).

When one varies the number of particles, we expect that the droplet expands without changing its plateaux density. We can see this to happen when $\beta=\frac{1}{2}$ in figure \two.

\begin{minipage}[!h]{\textwidth}
\framebox{\includegraphics[scale=0.3, angle=270]{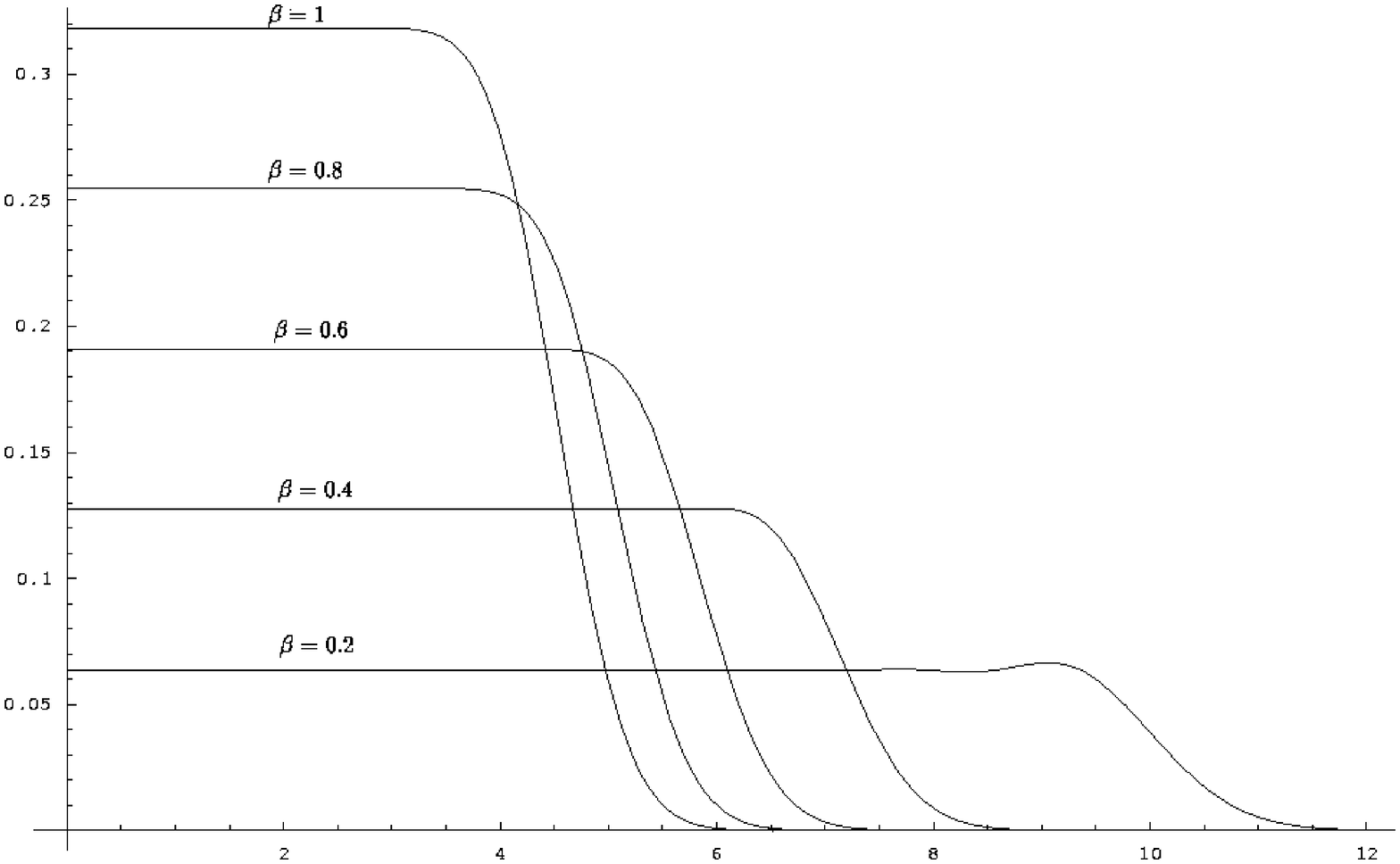}}
\framebox{\includegraphics[scale=0.31, angle=270]{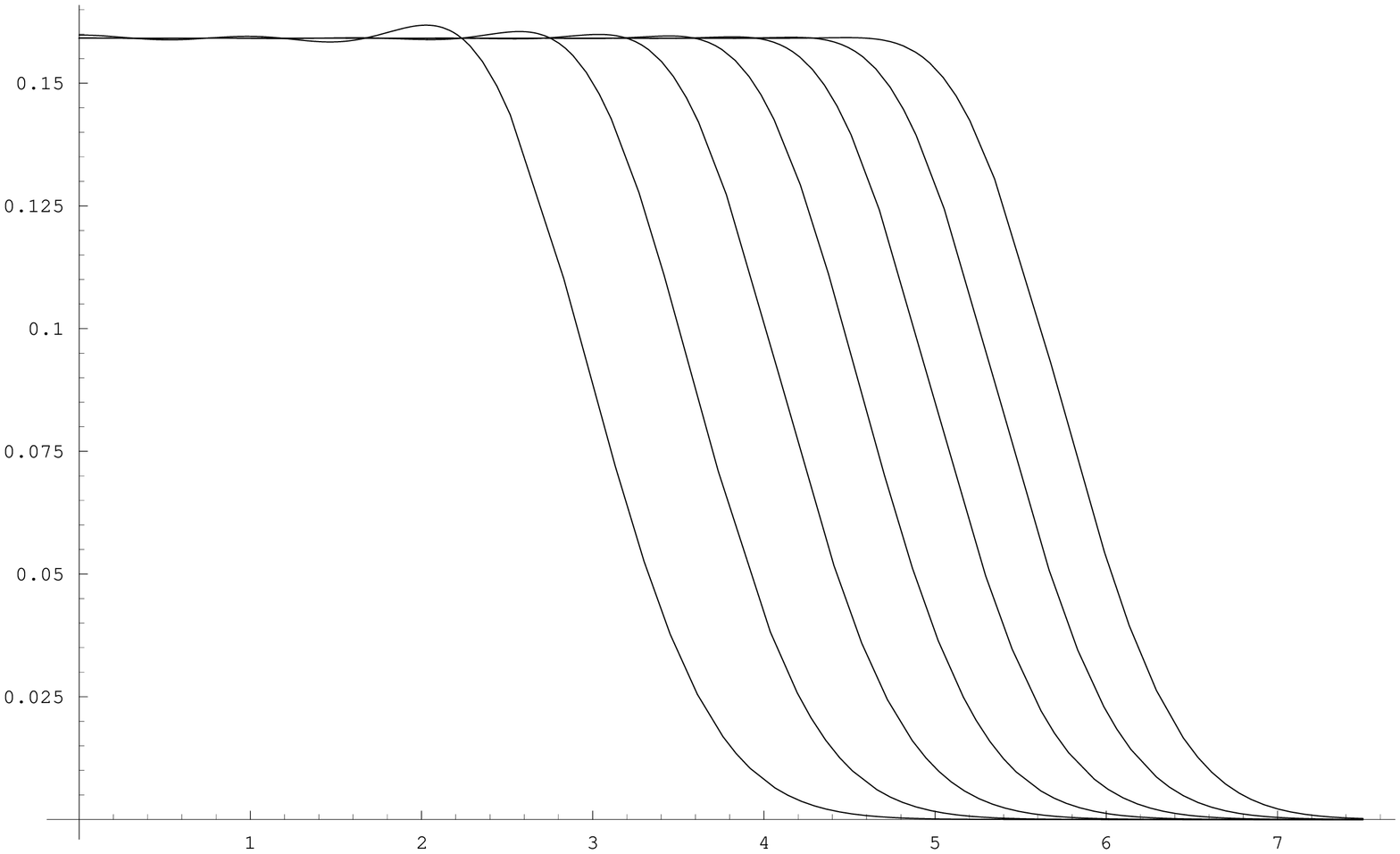}}
\begin{tabular}{p{210pt}p{210pt}}
\footnotesize{Figure \one: Density plot for various values of $\beta$.\phantom{asd}} &
\footnotesize{Figure \two: Density plot for different numbers of particles, $\beta=\frac{1}{2}$}
\end{tabular}
\end{minipage}\\

The exact expression of the density in $\eta=0$ is actually
\[
{}_{\beta}{\bra\Omega U[\check \rho(0,0)]\ket\Omega}_{\beta} = \frac{\beta}{\pi}\left[1-\left(\frac{\beta-1}{\beta+1}\right)^{N+1}\right] \longrightarrow \frac{\beta}{\pi} \ , \; {\rm for } \ N \gtrsim ({\rm a} \ {\rm few} \ {\rm units}).
\]
Yet, one can find this number in a simpler way. Take the \emph{square radius} operator $\hat R$
\[
\hat R^2 \doteq \frac{\{\hat z,\hbz\}_+}{2} = \frac{\{\hat b,\hat b^\dagger\}_+}{2\beta^2}+\frac{\{\hat a,\hat a^\dagger\}_+}{2}+\frac{\hat a\hat b+\hat a^\dagger \hat b^\dagger}{2\beta} \ ;
\]
the area of the droplet represented by $\ket\Omega$ is estimated by the expectation value of $\hat R^2$ on a Landau state in the lowest energy level
\begin{equation}
\bra{\psi_{N,0}} \hat R^2 \ket{\psi_{N,0}} = \frac{2N+\beta+1}{2\beta} = \frac{1+\beta}{2\beta}+\frac{N}{\beta} \ ,
\label{eqn:rsquare}
\end{equation}
and so the (average) density is estimated to be
\begin{equation}
\bar\varrho = \frac{N+1}{R^2 \pi} = \frac{2\beta}{\pi}\frac{N+1}{2N+\beta+1} \overset{\scriptscriptstyle{N\rightarrow\infty}}{\longrightarrow} \frac{\beta}{\pi} \ .
\label{eqn:estrho}
\end{equation}
We deduce that at large scale, in collective coordinates space (see footnote at page \pageref{collect}) the fluid $\ket\Omega$ is an incompressible fluid with filling factor rescaled by $\beta$ (see e.g. \cite{girvinprange}). From the second of the \pref{eqn:deform} we see that the flux of the magnetic field $\mathbf B$ through a surface of unit area is rescaled by the same factor of $\beta$, because of the noncommutativity. Notice that $\beta$ is a tunable parameter, hence the density of this fluid can mimic the fractional filling of the ordinary Landau levels.\\
This is, of course, only a statement regarding long distances physics, since such is the very definition of filling fraction. To go any further on the applicative side of the model, one should study the short distance physics as well. To get this done, one needs to study other correlation functions. This will be pursued in the next section.
\subsection{The correlation function $\langle\check \rho(x)\check \rho(y)\rangle$}
We turn back for a moment to $\beta=1$, in order to show the form of the density-density correlation function on the incompressible ground state $\bra\Omega\check \rho(z_1)\check \rho(z_2)\ket\Omega$. We will work out a form which holds also for $\beta\neq 1$,
and now try to compute the correlation function $\langle\check \rho(x)\check \rho(y)\rangle_\Omega$. A straightforward computation leads, in the same way as before, to the result 
\begin{eqnarray}
\fl\bra\Omega U[\check \rho (z_1,\bar z_1)] U[\check \rho(z_2,\bar z_2)] \ket\Omega = \nonumber\\
= \sum_{klmn} \, \psi^\ast_{l,0}(z_1,\bar z_1)\psi_{k,0}(z_1,\bar z_1)\psi^\ast_{m,0}(z_2,\bar z_2)\psi_{n,0}(z_2,\bar z_2)\,\bra 0 \check c_0\cdots \check c_N \, \check c^\dagger_l\check c_k\, \check c^\dagger_m\check c_n\,  \check c^\dagger_N\cdots \check c^\dagger_0\ket{0} = \nonumber\\
= \ \delta^2(z_1-z_2)\langle U[\check \rho(z_1,z_1)]\rangle \ + \nonumber\\
\ \ - \sum_{l\neq m}^N\scalprod{\psi_{l,0}}{\weyl{\delta(z_1-z)\delta(\bar z_1-\bar z)}\psi_{m,0}}\scalprod{\psi_{m,0}}{\weyl{\delta(z_2-z)\delta(\bar z_2-\bar z)}\psi_{l,0}} \ +\nonumber\\
\ \ + \sum_{l\neq m}^N\scalprod{\psi_{l,0}}{\weyl{\delta(z_1-z)\delta(\bar z_1-\bar z)}\psi_{l,0}}\scalprod{\psi_{m,0}}{\weyl{\delta(z_2-z)\delta(\bar z_2-\bar z)}\psi_{m,0}} \, .
\end{eqnarray}
This last expression is valid also for $\beta\neq 1$. Operating on it, one can see that the last two terms are
both real, and moreover they are both invariants under simultaneous rotations of $z_1$ and $z_2$ on the complex plane
\[
\left\{\begin{array}{ccl}
z_i & \mapsto & \rme^{i\phi}z_i \\
\bar z_i & \mapsto & \rme^{-i\phi}\bar z_i 
\end{array}\right. \ .
\]
We can considerably simplify the formula for the correlation function by computing it for $z_1=0$ and with $z_2$ on the real line $\eta = z_2 = \bar z_2$, away from the origin $\eta=0$. Using the \pref{eqn:deltas}, we obtain
\begin{eqnarray}
\fl\bra\Omega U[\check \rho(0)]U[\check \rho(\eta,\bar \eta)]\ket\Omega = & \frac{1}{\pi^2}\left|\frac{2\beta}{1+\beta}\right|^2\sum_{m=0}^N \,\left\{\frac{1+\beta}{2}\left(1-\left(\frac{\beta-1}{\beta+1}\right)^{N+1}\right)-\left(\frac{\beta-1}{\beta+1}\right)^m\right\}\times \nonumber \\
&\times \sum_{s=0}^{m}\binomial{m}{s}\left(\frac{2}{1+\beta}\right)^{m-s} \frac{\mathcal U(s-m,1,\frac{2\beta}{1+\beta} |\eta|^2)}{(m-s)!} \ \rme^{-\frac{2\beta}{1+\beta}|\eta|^2} \ .
\label{eqn:finalcorr}
\end{eqnarray}

The shape of the function as we vary the number of particles $N$, is left basically invariant within a characteristic length, the latter being basically the only object which varies with $N$. This is exactly the same as in the commutative case.\\[8pt]
\begin{minipage}[!h]{\textwidth}
\framebox{\includegraphics[scale=0.3125, angle=270]{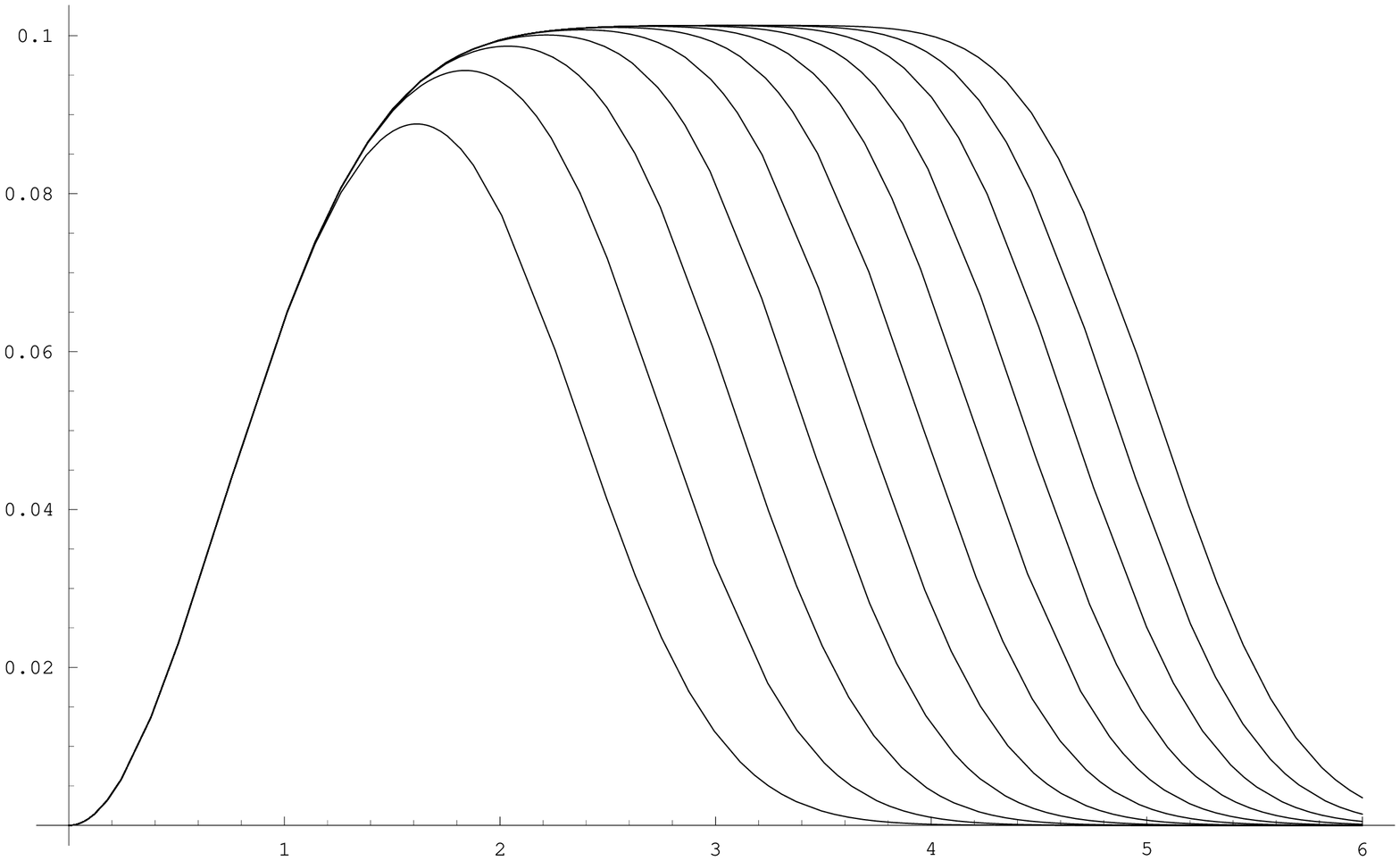}}
\framebox{\includegraphics[scale=0.52,angle=270]{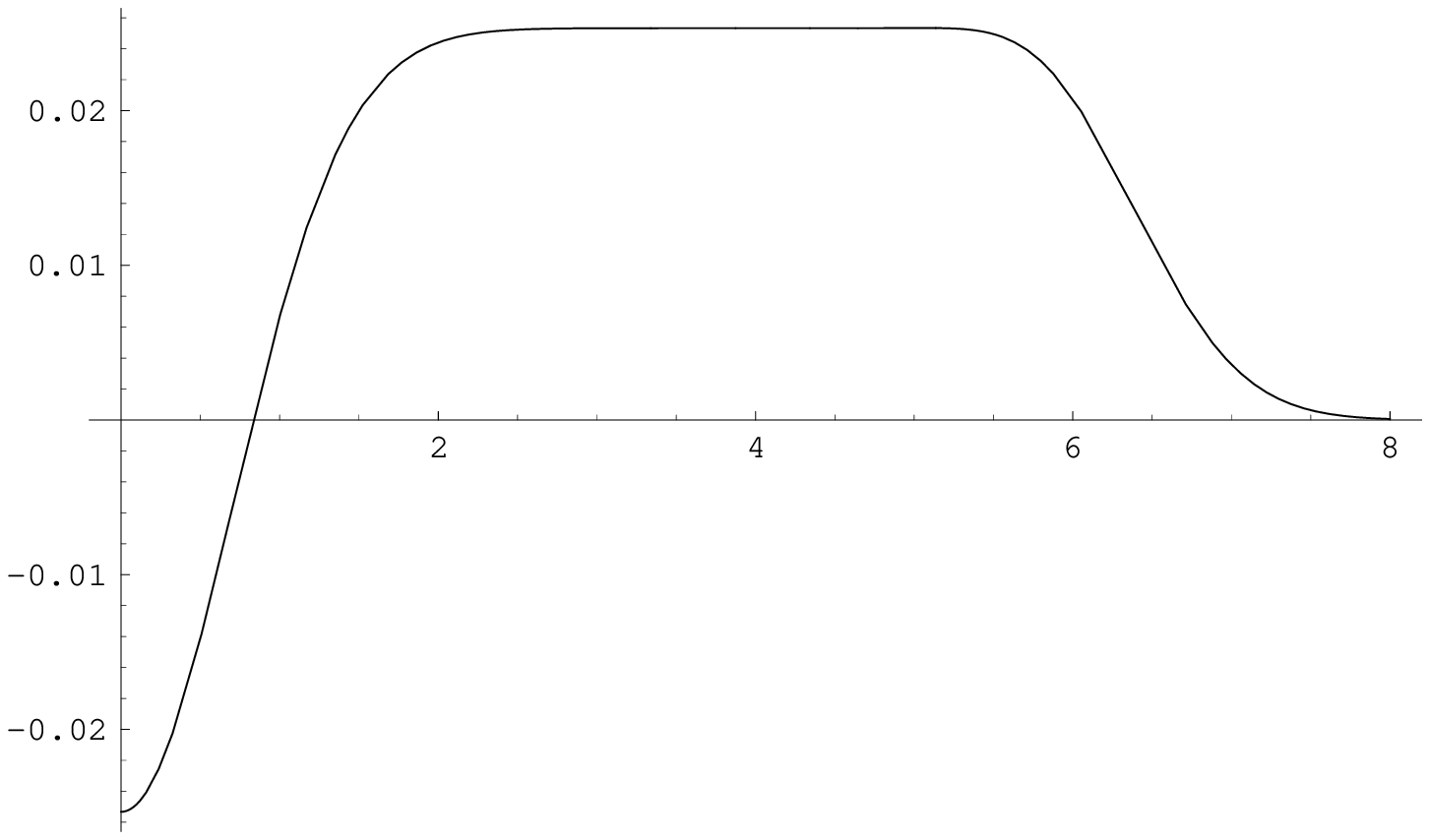}}
\begin{tabular}{p{210pt}p{210pt}}
\footnotesize{Figure \three: Plot of the correlation function of the density with itself for various numbers of particles for $\beta = 1$ (commutative case).} &
\footnotesize{Figure \five: Plot of the correlation function of the density with itself for $N=20$ particles, $\beta = \frac{1}{2}$.}\\
\end{tabular}
\end{minipage}\\

As it is apparent from figure 
\five and \three
, in the noncommutative case ($\beta\neq 1$) the two points correlation function of the density has an uncommon feature near the origin, because it becomes negative.\footnote{Anyhow, the quantity $\bra\Omega U[\rho(0)]U[\rho(0)]\ket\Omega$ is positive due to the contact term. Remember that in \pref{eqn:finalcorr} also exchange contributes are accounted for.} This is an effect of noncommutative deformation of the algebra of Landau levels. To understand it in physical terms, we can do the following: we switch on a small perturbation, in the form of a two body potential
\[
\hat V(x,y)\doteq\hat\psi(x)\hat\psi^\ast(x)V(x-y)\hat\psi(y)\hat\psi^\ast(y)
\]
and we compute the first order perturbation on the unperturbed ground state. The result is (for simplicity we do the computation at $x=0$, $y=\bar y = r$)
\[
\mathcal V(r) \doteq \mathcal V(0,r) = \bra\Omega U[\check \rho(0)] V(0,r) U[\check \rho(r)]\ket\Omega \ .
\]
In the case of the harmonic potential $V(0,r) = \frac{1}{2} r^2$, we obtain for the effective potential $\mathcal V(r)$ a shape which has a minimum at $r\neq 0$, as shown by figures 
\six \ and \seven.\\[8pt]
\begin{minipage}[!h]{\textwidth}
\framebox{\includegraphics[scale=0.3]{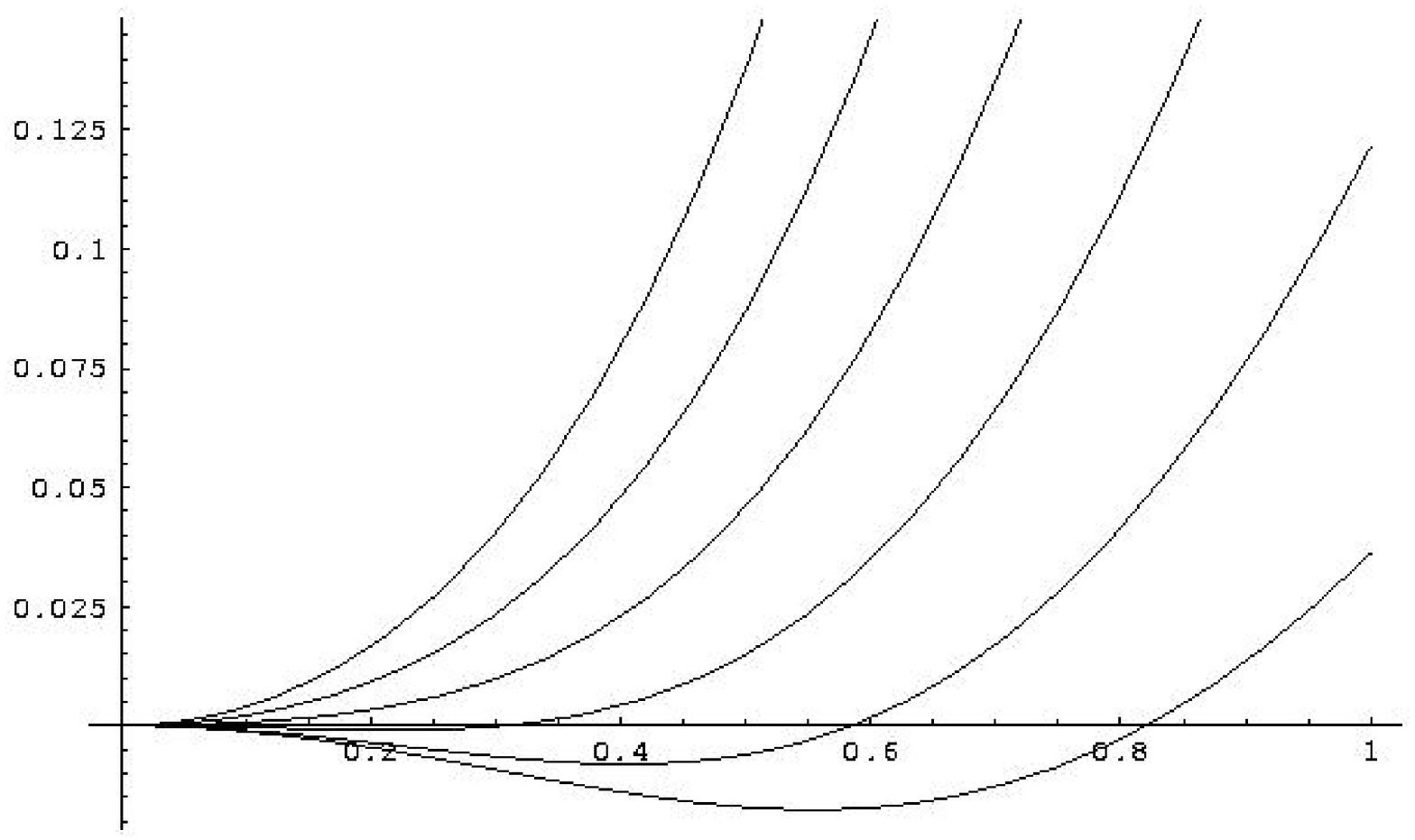}}
\framebox{\includegraphics[scale=0.343]{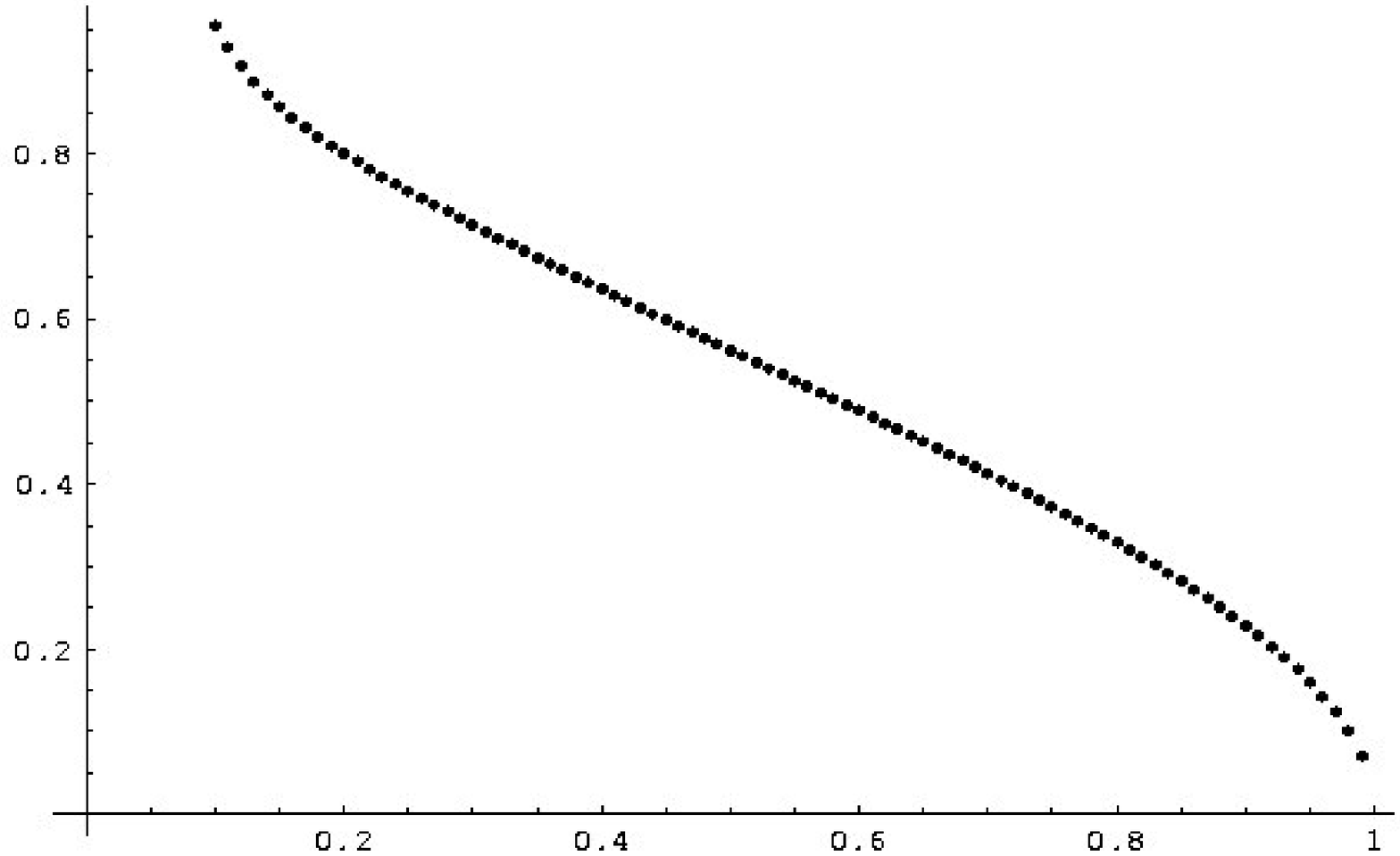}}
\begin{tabular}{p{210pt}p{210pt}}
\footnotesize{Figure \six: Effective potential for different values of $\beta$'s. \quad $\beta \in [0.5,1.0]$} & \footnotesize{Figure \seven: Locations of minima of the effective potential as a function of $\beta \in [0.1,1]$} \\
\end{tabular}
\end{minipage}\\

It means that the attraction between the particles due to $\hat V$ is balanced by an effective ``repulsion'' that is related to the loss of localization on the noncommutative plane (see also \cite{landincbook}). This effect, implying a sort of granularity of the incompressible fluid described by the formalism, has been used also in a recent description of Quantum Hall effect (see the papers \cite{QH-NCCS}, and references therein).

\section{Green function}
It is possible to define a noncommutative generalisation of the Green function of the usual Quantum Mechanics problem. One can define it for $\beta=1$ in the second quantization scheme, introducing also time dependence in field $\check \phi$ (e.g. \cite{landau9})
\begin{equation}
G(t,x;t',y) \doteq - \bra\Omega T(\check \phi(t;x,\bar x) \check \phi^\dagger(t';y,\bar y))\ket\Omega \ .
\end{equation}
Since the ground state $\ket\Omega$ is a product of energy-degenerate single-particle states, the time dependence is very simple. Indeed, besides contact terms, we have ($t<t'$)
\[
G(t,x;t',y) = \sum_{k=0}^N\psi^{\phantom{\ast}}_{k0}(x,\bar x)\psi^\ast_{k0}(y,\bar y) \rme^{-\rmi(t-t')} = G(x,y) \rme^{-\rmi(t-t')} \ .
\]
To generalise $G(x,y)$ to $\beta\neq 1$, we write
\begin{equation}
\hat\Gamma(x,y) \doteq \sum_{k=0}^N \, \weyl{\delta^2(x-\hat z)}|\psi_{k0})(\psi_{k0}|\weyl{\delta^2(y-\hat z)} \ .
\label{eqn:NCGreenf}
\end{equation}
The sum in the above expression comes from the evaluation of the average on the ground state $\bra\Omega$. We see indeed that in the $\beta\to 1$ limit, \pref{eqn:NCGreenf} becomes
\[
\hat\Gamma(x,y)\Big\arrowvert_{\beta=1} = \sum_{k=0}^{N} |x,\bar x) \ \psi^{\phantom{\ast}}_{k0}(x,\bar x)\psi^\ast_{k0}(y,\bar y) \ (y,\bar y|
\]
Now we can compute from \pref{eqn:NCGreenf} the function $\hat\Gamma(x,y)$ in the general case, by using the same techniques employed before, with a slightly harder computation (see appendix for a few details); in particular we compute the generic matrix element between the states $\psi_{lm}$ and $\psi_{l'm'}$ (now set $y=\bar y=0$ to display a simple expression)
\begin{eqnarray}
\fl(\psi_{lm}|\hat\Gamma(x,0)|\psi_{l'm'}) = & \left(\frac{1}{2\pi}\frac{2\beta}{1+\beta}\right)^2 \,  (-1)^{l'}\left(\frac{1-\beta}{2}\right)^{l'-m'} \, \sqrt{\frac{l!l'!}{m!m'!}} \ \chi^{[0,N]}(l'-m') \times \nonumber \\
& \times  \sum_{s=0}^{(l'-m')\curlywedge l}\frac{\beta^{s+m'-\frac{l+3l'}{2}}}{(l'-m'-s)!(l-s)!s!}\left(\frac{2\beta}{1+\beta}\right)^{m-m'+2l'-s} \times \nonumber \\
& \times \bar x^{m-m'+l'-l} \, \mathcal U(s-l,m-m'+l'-l+1,\frac{2\beta}{1+\beta}x\bar x) \rme^{-\frac{2\beta}{1+\beta}x\bar x} \ ,
\label{eqn:explNCGreenf}
\end{eqnarray}
where there appears the characteristic function
\[
\chi^{[0,N]}(x)=
\left\{\begin{array}{lcl}
= 1 & \implied & 0\leq x\leq N\\
= 0 & \implied & x \notin [0,N]
\end{array}\right. \ .
\]
One can compute the expectation value of any operator $\widehat{\bf F} $ that can be written as a sum of one particle operators, by using \pref{eqn:NCGreenf} or \pref{eqn:explNCGreenf} (notice that we factored out time dependence of Green function, so we keep track of the ordering, see \cite{landau9})
\[
\widehat{\bf F} = \sum_i \hat f^{(i)} \ \implies \ \langle \hat{\bf F} \rangle = \frac{1}{N} \int \, \rmd^2 x \ \tr \hat f^{(x)} \, \hat\Gamma(x,y) \Big\arrowvert_{x=y} \ .
\]
Further analysis of the observables of the noncommutative Landau levels will be discussed elsewhere.
\section{Conclusions}
The algebraic deformation performed in the present paper has shown itself to have interesting physical features. First of all, it has been shown that, owing to the preservation of the $W_\infty$-algebra, it still makes sense to speak about incompressibility. Hence the completely filled state $\ket\Omega$ has been shown to be still incompressible after the deformation. It has been displayed a technique for dealing with this Quantum Mechanics on Noncommutative Plane, which relies on the operatorial generalisation of Dirac $\delta$-function (see section \ref{sec:weyltransform}); as a consequence, density and Green function on the ground state have been computed: in particular the two-point correlation function of density gives us an important key for understanding the physical precise meaning of the delocalization generated by noncommutativity, which has been shown to affect the short distance behaviour of the fluid itself. Also, the long range effect has been shown to be the rescaling of the flux of magnetic field $\mathbf B$, and consequently the rescaling of filling fraction of the ground state. More extended analysis of this problem are deferred to a following work, in particular the application to the theory of the Fractional Quantum Hall effect.
\ack
I would like to thank A.Cappelli for discussions and collaboration in the early stages of this work. I also thank D.Seminara for useful discussions.\\
This work has been partially funded by the EC Network contract HPRN-CT-2002-00325, {\em ``Integrable models and applications: from strings to condensed matter''}.

\appendix
\section{The computation of the Green function }
To compute the explicit expression of \pref{eqn:NCGreenf}, we need the quantity
\begin{eqnarray}
\fl\scalprod{\psi_{l,m}}{\weyl{\delta(x-\hat z)\delta(\bar x-\hbz)}\psi_{k,0}} = \int \frac{\rmd x\rmd y}{(2\pi)^2}\, \rme^{\rmi(qx+py)} \scalprod{\rme^{\frac{\rmi}{\beta}\bar y\hat b^\dagger}\rme^{\rmi\bar y\hat a}\psi_{l,m}}{\rme^{-\frac{\rmi}{\beta}x\hat b^\dagger}\psi_{k,0}} \rme^{\frac{1-\beta}{2\beta}xy} \, .
\end{eqnarray}
Now we can use the fact that
\[
\scalprod{\rme^{\frac{\rmi}{\beta}\bar y\hat b^\dagger}\rme^{\rmi\bar y\hat a}\psi_{l,m}}{\rme^{-\frac{\rmi}{\beta}x\hat b^\dagger}\psi_{k,0}} = (-1)^k\mathcal U\left(-k,1,xy\right)\scalprod{\rme^{\frac{\rmi}{\beta}\bar y\hat b^\dagger}\psi_{l,0}}{\rme^{-\frac{\rmi}{\beta}x\hat b^\dagger}\psi_{k,0}}
\]
and equation \pref{eqn:shortmatel}, to find that
\begin{eqnarray}
\fl\scalprod{\psi_{l,m}}{\weyl{\delta(x-\hat z)\delta(\bar x-\hbz)}\psi_{k,0}} = \frac{1}{2\pi} & \left|\frac{2\beta}{1+\beta}\right|\sqrt{\frac{k!l!}{m!}} \, \sum_{s=0}^{k\curlywedge l}\frac{\beta^{s-\frac{k+l}{2}}}{(k-s)!(l-s)!s!} \, \left(\frac{2\beta}{1+\beta}\right)^{m+k-s}\times \nonumber \\
& \times(\bar x)^{m+k-l}\, \mathcal U\big(s-l,m+k-l+1,\frac{2\beta}{1+\beta}x\bar x\big) \rme^{-\frac{2\beta}{1+\beta}x\bar x} \, .
\label{eqn:firststep}
\end{eqnarray}
We can compute the matrix elements of $\hat\Gamma(x,y)$ using the above, indeed we have
\begin{eqnarray}
\fl(\psi_{lm}|\hat\Gamma(x,y)|\psi_{l'm'}) = \sum_{k=0}^N \, \scalprod{\psi_{lm}}{\weyl{\delta^2(x-\hat z)} \psi_{k0}}\scalprod{\psi_{k0}}{\weyl{\delta^2(y-\hat z)}\psi_{l'm'}} \, .
\label{eqn:rawmatel}
\end{eqnarray}
Now we can insert the \pref{eqn:firststep} into the \pref{eqn:rawmatel}, noticing that
\[
\scalprod{\psi_{l,m}}{\weyl{\delta(x-\hat z)\delta(\bar x-\hbz)}\psi_{k,0}} = \scalprod{\psi_{k,0}}{\weyl{\delta(x-\hat z)\delta(\bar x-\hbz)}\psi_{l,m}}^\ast \, . 
\]
Putting $y=0$ we find the simplified expression \pref{eqn:explNCGreenf}. Of course to compute expectation values we need the more complicated matrix elements with $x=y$ (see the text).


\section*{References}
\begin{thebibliography}{99}
\bibitem{fubini} S.Fubini, {\bf Mod.Phys.Lett. A}6, (1991), p.347.
\bibitem{fradkinbook} E.Fradkin, {\it ``Field Theory of Condensed Matter systems''}, Addison-Wesley, (1991), New York.
\bibitem{girvinprange} S.M.Girvin,~R.E.Prange, {\it "The Quantum Hall Effect"}, Graduate Texts in Contemporary Physics, Springer Verlag, (1990), New York.
\bibitem{infsym} A.Cappelli,~C.A.Trugenberger,~G.R.Zemba, {\it ``Infinite Symmetry in the Quantum Hall effect''}, {\bf Nucl.Phys. B}396, (1993), p. 465-490; {\tt hep-th/9206027}
\bibitem{qmonncplandsph} V.P.Nair,~A.P.Polychronakos, {\it ``Quantum Mechanics on the Noncommutative Plane and Sphere''}, {\bf Phys.Lett. B}505, (2001), p. 267-274; {\tt hep-th/0011172}. {\it See also:} C.Duval,~P.A.Horvathy, {\it ``The "Peierls substitution" and the exotic Galilei group''}, {\bf Phys. Lett. B}479, (2000), 284 {\tt hep-th/0002233}. ~C.Duval,~P.A.Horvathy, {\it ``Exotic Galilean symmetry in the non-commutative plane, and the Hall effect.''}, {\bf J.\ Phys.\ A}34 (2001), 10097; {\tt hep-th/0106089}.
\bibitem{landau9} L.D.Landau,~E.M.Lifschitz, {\it ``Course of Theoretical Physics: Statistical Physics part II''}, Pergamon (1980), Oxford.
\bibitem{landincbook} G.Landi, {\it ``An Introduction to Noncommutative spaces and their Geometry''}, Lecture Notes in Physics, New Series M, Springer Verlag, (1998); \emph{also:} {\tt hep-th/9701078} 
\bibitem{QH-NCCS} L.Susskind, {\it ``The quantum Hall fluid and non-commutative Chern Simons theory.''}, {\tt hep-th/0101029.} \\ A.P.Polychronakos, {\it ``Quantum Hall states as matrix Chern-Simons theory.''}, {\bf JHEP} 0104, (2001),p. 011, {\tt hep-th/0103013.} \\ A.Cappelli,~M.Riccardi, {\it ``Matrix model description of Laughlin Hall states.''}, \mbox{\bf J. Stat. Mech.}, (2005) P05001, {\tt hep-th/0410151}.
\bibitem{spallucci} A.Smailagic,~E.Spallucci, {\it ``Feynman path integral on the noncommutative plane.''}, {\bf J.Phys. A}36, (2003) L467; {\tt hep-th/0307217}.
\bibitem{handbook} M.Abramowitz,~I.A.Stegun~(Eds.), {\it "Confluent Hypergeometric Functions."} Ch. 13 in {\it ``Handbook of Mathematical Functions with Formulas, Graphs, and Mathematical Tables''}, 9th printing. New York: Dover, (1972) pp. 503-515.
\end{thebibliography}
\end{document}